\newcount\eqnno \eqnno=0 
\def\numeqn{\global\advance\eqnno by 1 \eqno(\the\eqnno)} 
\def\etal{et al. }
\def\gsim{{ _ >\atop{^\sim}}}  
\def\lsim{{ _ <\atop{^\sim}}}

\documentstyle{mn}
\input psfig

\title[The non-linear correlation function]{The non-linear correlation 
function and the shapes of virialized halos}

\author[Ravi K. Sheth and Bhuvnesh Jain]
{Ravi K. Sheth$^1$ and Bhuvnesh Jain$^2$\\
$^1$ Berkeley Astronomy Department, University of California, 
Berkeley, CA  94720\\
$^2$ Max-Planck-Institut f\"ur Astrophysik,
Karl-Schwarzschild-Strasse 1, 85748 Garching, Germany\\
\smallskip
Email: sheth@astron.berkeley.edu, bjain@mpa-garching.mpg.de\\
}
      
\begin{document}

\maketitle

\begin{abstract}
The correlation function $\xi(r)$ of matter in the non-linear regime is 
assumed to be determined by the density profiles $\rho(r)$ and the mass 
distribution $n(M)$ of virialized halos. The Press--Schechter approach 
is used to compute $n(M)$, and the stable clustering hypothesis is used to 
determine the density profiles of these Press--Schechter halos. Thus, 
the shape and amplitude of $\xi(r)$ on small scales is related to the 
initial power spectrum of density fluctuations.

The case of clustering from scale-free initial conditions is treated in
detail.  If $n$ is the slope of the initial power spectrum of density 
fluctuations, then stable clustering requires that 
$\xi(r)\propto r^{-\gamma}$, where $\gamma$ is a known function of
$n$.  If halo--halo correlations can be neglected, then 
$\rho(r)\propto r^{-\epsilon}$, where 
$\epsilon = (\gamma+3)/2 = 3(4+n)/(5+n)$.  For all values of $n$ of
current interest, this slope is steeper than the value $3(3+n)/(4+n)$ 
that was obtained by Hoffman \& Shaham in their treatment of the shapes 
of the outer regions of collapsed halos.  
Our main result is a prediction for the amplitude of the non-linear 
correlation function. The predicted amplitude and its dependence on $n$ 
are in good quantitative agreement with $N$-body simulations of 
self-similar clustering.  

If stable clustering is a good approximation only inside the half-mass
radii of Press--Schechter halos, then the density contrast required 
for the onset of stable clustering can be estimated. This
density contrast is in the range $\sim 300-600$ and 
increases with the initial slope $n$, 
in agreement with estimates from $N$-body simulations. 

\end{abstract} 
\maketitle

\section{Introduction}

Large-scale structure in the universe is thought to arise through
the gravitational clustering of matter. In the nonlinear regime most
of the dark matter is in bound halos which have separated out from
the expanding background. Therefore the auto-correlation function, 
$\xi(r)$, of the dark matter is closely related to the shapes of halos.
In this paper we explore the consequences of this connection
for the shape and amplitude of $\xi(r)$ on small scales. 

There are two approximations that are commonly used to describe the 
evolution of gravitational clustering in the non-linear
regime.  The first is the assumption that, in this regime, the clustering 
is statistically stable.  By this, one usually means that non-linear, 
virialized objects no longer participate in the expansion of the background 
Universe; they maintain their shapes in physical coordinates, so they 
shrink in comoving coordinates (e.g. Peebles 1965, 1980).  The second 
approximation was first formulated by Press \& Schechter (1974).  They 
assumed that, on average, structures collapse spherically and then 
virialize, with initially more dense regions collapsing first, and less dense 
regions collapsing later.  Further, they assumed that when they virialize, 
all halos have the same density (about 200 times the background density 
at the time of virialization) whatever their mass.  By applying these
assumptions to an initially Gaussian density field, they derived the
distribution of halos as a function of mass.  In the Press--Schechter 
approach, the clustering is hierarchical, and virialized halos at a 
given epoch are progenitors of the halos that virialize at a later epoch.  
It is not obvious that these two idealizations, stable clustering, and 
virialization at a fixed multiple of the background density at the time of 
virialization, are compatible.  

For example, consider a Press--Schechter halo with uniform density (it 
has a `tophat' density profile).  Assume that at the time it virializes, 
say $t_1$, it has mass $M_1$ and density $178\rho_{\rm b}(t_1)$.  This 
sets its radius $R_1$.  In the Press--Schechter approach, at some later 
time $t_2>t_1$, it will have merged with other halos into an object of 
greater mass $M_2>M_1$.  This more massive object will have density 
$178\,\rho_{\rm b}(t_2)<178\,\rho_{\rm b}(t_1)$.  On the other hand, 
if stable clustering is correct, then the core $M_1$ within $R_1$ will 
still have density $178\,\rho_{\rm b}(t_1)$ so that the shell of mass 
$(M_2-M_1)$ around the core must be less dense than the core itself.  Thus, 
if it had a tophat density profile at time $t_1$, then the halo will not 
have a tophat density profile at the epoch $t_2$.  In other words, if 
Press--Schechter halos are to evolve consistently with the stable 
clustering assumption, then they cannot have tophat density profiles.  
This toy example illustrates that by requiring consistency between 
Press--Schechter and the stability approximation we can constrain the 
allowed density profiles of halos.  

In particular, it is known that the stable clustering approximation 
allows one to constrain the shape of $\xi(r)$ in the non-linear regime 
(e.g., Peebles 1974a; section 26 in Peebles 1980; section 5.4 in 
Padmanabhan 1993). Section 2 uses this stable clustering 
shape of $\xi(r)$ to constrain the density profiles of Press--Schechter 
halos. The formalism of McClelland and Silk (1977) is then used to
compute the amplitude of $\xi(r)$ in the nonlinear regime. The
predicted amplitude is compared with the results of N-body
simulations. The density profile inferred in Section~2 is different
from the value that is obtained in other descriptions of spherical 
collapse, such as the Hoffman-Shaham model.  Section~3 discusses 
the reasons for the difference.  Section~4 studies the density contrast 
required for the onset of stable clustering.  It argues that only the 
cores of Press--Schechter halos are likely to have evolved consistently 
with both Press--Schechter and stable clustering.  This has the 
consequence that the density required for the onset of stable 
clustering should be an increasing function of $n$, where $n$ is the 
slope of the power spectrum of initial fluctuations.  Section~5 
discusses the case of non-power law density profiles and summarizes 
the results. 

\section{The correlation function in the non-linear regime}

In the highly non-linear regime, stable clustering may be a good 
approximation.  This section combines the stable clustering 
hypothesis with the Press--Schechter model to compute the shape 
and amplitude of the correlation function $\xi(r)$.  
We consider first some simple and general properties of $\xi$. 

\subsection{Stable clustering and the density profile}
Consider the auto-correlation function of matter $\,\xi(r)\,$ within a 
single halo. If the density profile of the halo is a power law,
\begin{equation}
\rho(r)=A r^{-\epsilon} ,
\label{profile}\end{equation}
then $\,\xi(r)$ is
\begin{eqnarray}
\xi(r) \, \propto\int {\rm d}^3s \,\rho(\vec s)\, \rho(\vec r + \vec s)\,\,
\propto \int {\rm d}s\, s^2 \, s^{-\epsilon} \,\, |\vec r +\vec
s|^{-\epsilon}\, 
\label{convo}
\end{eqnarray}
which is just a convolution of the density profile with itself.
When the initial profile satisfies $3/2<\epsilon<3$, then the 
integral converges to give a power law form for the correlation
function (Peebles 1974b; McClelland \& Silk 1977):
\begin{equation}
\xi(r)\propto r^{-\gamma} \, , \qquad {\rm where}\,\, \gamma=2\epsilon-3 \, .
\label{gamma-eps}
\end{equation}
If the density profile is a power law with slope $\epsilon$ on small scales, 
but has a cutoff on large scales, then the correlation function is also a 
power law, with slope $\gamma$ given by equation~(\ref{gamma-eps}) on small 
scales, but with a cutoff on larger scales.  

Assuming that all matter is in halos of various masses, 
$\xi(r)$ depends on correlations within halos
(essentially the density profile), correlations between halos, and the 
distribution of halo masses and radii.  McClelland \& Silk (1977) show
that if all halos have the same mass and radius, and these halos are 
randomly placed in space (so that there are no halo--halo correlations), 
then the above result (equation~\ref{gamma-eps}) still relates the shape 
of $\xi$ to the density profile $\rho$.  They also show that, for randomly 
placed halos having a range of different masses, but the same density profile,
an integral over mass must be included.  This integral affects the 
amplitude, but not the shape of $\xi$.  (If the density profile has a mass 
dependent cutoff, then the position of the `knee' in $\xi$ will depend 
on the number density of halos as a function of halo mass.)  

Of course, if the halos are not randomly distributed, then their 
correlations will change the shape of $\xi$.  Since halos are likely to
be correlated with each other, the relation (equation~\ref{gamma-eps}) 
between the shapes of $\xi$ and $\rho$ would seem to be of limited 
applicability.  However, if we restrict attention to small scales 
(the highly non-linear regime), then, for sufficiently small separations $r$, 
both members of each pair of particles will almost certainly be drawn 
from the same halo.  This means that, for small separations, the shape 
of the correlation function is not affected by halo--halo correlations.  
If the halo--halo correlation function can be neglected, then we can use 
the formalism of McClelland \& Silk (1977) to compute the shape and 
amplitude of $\xi(r)$ in the non-linear regime. To do so, however, we must 
know the density profiles of halos, and the number density of halos as a 
function of halo mass.  

In the highly non-linear regime, stable clustering is a good approximation, 
and it can be used to constrain the shapes of virialized halos.  
For clustering from scale-free initial conditions,  
Davis \& Peebles (1977) show that the BBGKY hierarchy of equations
admit self-similar solutions (Peebles 1980; Section 73). 
They show that, in the stable clustering regime, $\xi(r)$ 
must be a power law:  
$\xi(r)\propto r^{-\gamma}$ with $\gamma = 3(3+n)/(5+n)$, where $n$ is the 
slope of the initial spectrum $P(k)\propto k^n$.  This form for $\xi(r)$,
combined with equation~(\ref{gamma-eps}), yields a value for $\epsilon$ in 
the stable clustering regime: 
\begin{equation}
\epsilon_{\rm sc} = {\gamma + 3\over 2} = {3(4+n)\over (5+n)} ,\qquad 
{\rm for}\ \ n>-3\, .
\label{sc} 
\end{equation}

Note that whereas $\epsilon_{\rm sc}$ may 
be thought of as the slope of the two-point correlation function, subject 
to the constraint that one member of each pair of particles is certainly 
at the center of a halo, $\gamma$ is the slope of the correlation 
function when the locations of the members of each pair are 
unconstrained.  For $0<\gamma<3$, equation~(\ref{sc}) shows that 
$\epsilon>\gamma$, which means that the correlation function centered 
on peaks is steeper than the unconstrained correlation function.  This 
behaviour is similar to that of peaks associated with Gaussian random 
fields; the density profile of a typical spherically averaged Gaussian 
peak is steeper than the unconstrained correlation function of the 
Gaussian field (Bardeen \etal 1986).

In the remainder of this paper we will use this density profile
$\rho(r)\propto r^{-\epsilon_{\rm sc}}$ to describe the shapes of 
virialized halos.  The differences between this profile and that 
suggested by Hoffman \& Shaham (1985) are discussed in Section~3.  The 
effects of non-power law density profiles such as that proposed by 
Navarro, Frenk \& White (1995) will be considered in Section~5.  

\subsection{The correlation function}

The previous subsection argued that the McClelland \& Silk formalism can
be used to compute to shape and amplitude of the correlation function
in the non-linear regime.  To do so requires knowledge of the density
profiles of halos, as well as the number density of halos as a function 
of mass.  It argued that the stable clustering hypothesis could be used to
determine the density profiles of the halos.  This section uses the
Press \& Schechter (1974) model to determine the mass function.  Then, it
combines the stable clustering density profile with the Press--Schechter 
mass function to compute the amplitude of the matter correlation function 
$\xi(r)$ in the non-linear regime.  

In the stable clustering regime, we can use the McClelland \& Silk formalism
to express $\xi(r)$ as 
\begin{equation}
\xi(r) = {\mu\int\!\!\int M^2\,
\lambda_\sigma (r)\,p(M,\sigma)\,{\rm d}M\,{\rm d}\sigma \over 
\left[\mu\int\!\!\int M\,p(M,\sigma)\,{\rm d}M\,{\rm d}\sigma\right]^2} ,
\label{mcsilk}
\end{equation}
where $\mu$ is the total number density of halos, 
$p(M,\sigma)\,{\rm d}M\,{\rm d}\sigma$ is the probability that a halo 
has mass in the range ${\rm d}M$ about $M$ and radius in the range 
${\rm d}\sigma$ about $\sigma$, and 
\begin{equation}
\lambda_\sigma({\bmath r}) = \int u_\sigma({\bmath s})\,
u_\sigma({\bmath s+r})\,{\rm d}{\bmath s},
\label{lambdar}
\end{equation}  
where $u_\sigma$ describes the average density profile of a halo of 
radius $\sigma$ (note the similarity to equation~\ref{convo}).  Here, 
$\int_0^\sigma M\,u_{\rm \sigma}(r)\,{\rm d}{\bmath r} = M$, 
where $M$ is the total mass of the halo.  Thus, in this regime, 
$\xi(r)$ is determined solely by the distribution of masses, radii, shapes, 
and density profiles of halos.  Below, we will combine the Press--Schechter 
model with the stable clustering hypothesis to specify these distributions.  

In the Press--Schechter description of non-linear clustering, all matter 
is assumed to be contained within spherical virialized halos, and the 
distribution of halo masses is determined by the initial conditions.  
For a scale free initial spectrum $P(k)\propto k^n$
and $\Omega = 1$, the number 
density of halos with mass in the range ${\rm d}M$ about $M$ is 
\begin{equation}
n(M)\,{\rm d}M\! =\! {\bar\rho\over \sqrt{\pi}} 
\left({M\over M_*}\right)^{{(n+3)\over 6}}\!
{\rm e}^{-\left({M\over M_*}\right)^{{(n+3)\over 3}}}\!
\left({n+3\over 3}\right){{\rm d}M\over M^2}
\label{nmps}
\end{equation}
(Press \& Schechter 1974; Lacey \& Cole 1993).  Here 
\begin{equation}
M_*\propto a^{6/(n+3)}
\label{mstar}
\end{equation} 
is a characteristic mass which grows as the universe expands.  This 
characteristic mass defines two scales.  The first scale, which we will
denote $r_0$, is defined as follows.  If the variance in the mass contained 
in randomly placed cells of radius $r_0$ is $1.68^2/2$, then the mean mass 
contained in these spheres is $M_* = (4\pi r_0^3/3)\,\bar\rho$, where 
$\bar\rho\propto a^{-3}$ is the background density (e.g. Efstathiou 
\etal 1988).  The second scale, denoted $r_*$, is the 
virial radius of an $M_*$ halo, and is defined below.  

The average density within each virialized halo is also known, so 
that the mass and radius of a virialized halo are related.  For the 
halos described by equation~(\ref{nmps}) above, 
$M=(4\pi\sigma^3/3)\,\Delta_{\rm nl}\bar\rho$, where 
$\Delta_{\rm nl} = 178$, and $\bar\rho$ is the 
background density as before.  Thus, in the Press--Schechter model, the 
distribution of halo masses, radii, and shapes are specified, 
whereas the density profiles of these halos are not.  So, in the 
Press--Schechter approach, the correlation function in the non-linear 
regime is 
\begin{equation}
\xi(r) = {\int M^2\,\lambda_M(r)\,n(M)\,{\rm d}M \over 
\left[\int M\,n(M)\,{\rm d}M \right]^2},
\label{mclps}
\end{equation}
where $n(M)\,{\rm d}M$ is given by equation~(\ref{nmps}).  Notice that 
equation~(\ref{nmps}) implies that the denominator in 
equation~(\ref{mclps}) is just $\bar\rho^2$.  Moreover, notice that
since $\lambda(r)$ (equation~\ref{lambdar}) is an expression like 
equation~(\ref{convo}), equation~(\ref{mclps}) is exactly of the form 
discussed in the previous subsection.  That is, it is a sum over all the 
different masses of terms like equation~(\ref{convo}).  This means that 
the shapes of the correlation function and the density profile are 
related to each other (equation~\ref{gamma-eps}).  

To proceed, we need expressions for $\lambda_M(r)$, which in turn depend 
on the density profiles $u_\sigma$ of the Press--Schechter halos.  If we 
assume that these density profiles are power laws in distance from 
the halo center, then stable clustering requires that the density profile 
must be $u_\sigma \propto r^{-\epsilon}$ with $\epsilon$ given by 
equation~(\ref{sc}).  Thus, the Press--Schechter model specifies the number 
density of halos as a function of mass, as well as providing a relation 
between the mass and radius of each halo, and, in the non-linear regime, the 
stable clustering hypothesis specifies the density profiles of the 
Press--Schechter halos.  So, by combining the Press--Schechter 
model with the stable clustering hypothesis, we are able to compute the 
shape and amplitude of $\xi(r)$.  

If we use the stable clustering value, $\epsilon = 3(4+n)/(5+n)$, for 
the slope of the density profile of Press--Schechter halos, then we 
can compute $\lambda_M(r)$.  For $-2<n\le 1$, we know that 
$\epsilon>2$, so   
\begin{eqnarray}
\lambda_M(r) &=& \left({3-\epsilon\over 4\pi\sigma^3}\right)^2 
{4\pi\sigma^{2\epsilon}\over r(\epsilon -2)} \nonumber \\
&&\qquad \times \quad \int_{r/2}^\sigma 
\left[|s-r|^{2-\epsilon} - s^{2-\epsilon}\right]\, s^{1-\epsilon}\,{\rm d}s ,
\end{eqnarray}
for all $0\le r\le 2\sigma$, and $\lambda_M(r) = 0$ otherwise.  
Similarly, when $n=-2$, so $\epsilon = 2$, then 
\begin{equation}
\lambda_M(r) = {1\over 4\pi\sigma^2r} \int_{r/2}^\sigma 
\ln \Bigl\vert{s\over r-s}\Bigr\vert\, {{\rm d}s\over s}\qquad 
{\rm if}\ 0\le r\le 2\sigma ,
\end{equation}
and $\lambda_M(r) = 0$ otherwise.    

In the limit of vanishingly small separations, 
\begin{eqnarray}
\lambda_M(r)&=&{1\over 4\pi\sigma^3}\left({\sigma\over r}\right)^2,\qquad \ \ 
\epsilon = 5/2, \ n=1 , \\
\lambda_M(r)&=&{1.22\over 4\pi\sigma^3}\left({\sigma\over r}\right)^{1.8},
\qquad \epsilon = 2.4, \ \ n=0 , \\
\lambda_M(r)&=&{1.58\over 4\pi\sigma^3}\left({\sigma\over r}\right)^{1.5},
\qquad \epsilon = 9/4, \ \,n=-1 , \\
\lambda_M(r)&=&{\pi^2/4\over 4\pi\sigma^3}\left({\sigma\over r}\right),
\qquad \ \ \ \,\epsilon = 2, \ \ \ \ \,n=-2 .
\end{eqnarray} 
If we absorb all constants into one term, then we can write 
$\lambda_M(r) = \lambda_n\sigma^{\gamma-3}/r^\gamma$, so that the 
correlation function is
\begin{eqnarray}
\xi(r) &=& {\lambda_n\over r^{\gamma}\bar\rho^2}
\left({4\pi\Delta_{\rm nl}\bar\rho\over 3}\right)^{{2\over (5+n)}} 
\int M^{{(3+n)\over(5+n)}}\,M\,n(M)\,{\rm d}M \nonumber \\
&=& {\lambda_n\over r^{\gamma}\bar\rho}
\left({4\pi\Delta_{\rm nl}\bar\rho\over 3}\right)^{{2\over (5+n)}} 
{M_*^{{(3+n)\over(5+n)}}\over \sqrt{\pi}}\ 
\Gamma\!\left({11+n\over 10+2n}\right), \nonumber \\
&=& \left({r_*\over r}\right)^{\gamma}\ 
\left({\lambda_n\over \sqrt{\pi}}\,{4\pi\Delta_{\rm nl}\over 3}\right)\ 
\Gamma\!\left({11+n\over 10+2n}\right),
\label{xips}
\end{eqnarray}
where we have used the fact that $4\pi\Delta_{\rm nl}\bar\rho\sigma^3/3=M$.  
(Notice that this means that $r_*^3=3M_*/4\pi\Delta_{\rm nl}\bar\rho$, so 
that $r_*$ is quite a lot smaller than the scale $r_0$ on which the variance 
is $1.68^2/2$.)  This, then, is the correlation function in the highly 
non-linear, stable clustering regime.  

It is interesting to compare this result with that which we would have
obtained had all halos had the same mass, say $M_{\rm halo} = f^3\,M_*$.  
In this case, equation~(\ref{mclps}) shows that the correlation function 
is just $\xi(r) = \lambda_n\,(4\pi\Delta_{\rm nl}/3)\,(f\,r_*/r)^\gamma$, for 
$r<r_{\rm halo}$, so that the amplitudes are comparable if 
$f^{\gamma} = \Gamma(11+n/10+2n)/\Gamma(1/2)$.  Note that the amplitude 
increases as $f$ increases, and is comparable to the Press--Schechter 
amplitude when $M_{\rm halo}$ is considerably less than $M_*$ ($0.42$ and 
$0.125$ times $M_*$, for $n=1$ and $n=-2$, respectively).  

\subsection{Comparison with simulations}

To illustrate the accuracy of the amplitude predicted by 
equation~(\ref{xips}), we can compare it directly with that measured 
in the $N$-body simulations of clustering from scale free initial 
conditions.  A simple and reliable way to make the comparison is to
use recently proposed fitting formulae for the non-linear $\xi$ which
have been calibrated using high resolution $N$-body simulations. 
These are based on the ansatz of Hamilton \etal (1991) which 
provides a universal relation between the linear and
non-linear correlation function (also see Gott \& Rees 1975). This 
ansatz has recently been extended and refined
(Nityananda \& Padmanabhan 1994; Peacock \& Dodds 1994; Padmanabhan 
\etal 1995; Padmanabhan 1995; Jain, Mo \& White 1995 -- JMW).  
We shall use the fitting formulae proposed by JMW which take into 
account a simple dependence on the shape of the initial spectrum that 
is exhibited by the $N$-body data. 

In the asymptotic non-linear regime the formulae are consistent
with stable clustering and take the form (see equations (5a) and (6a) 
of JMW),
\begin{equation}
\bar\xi_{\rm NL}(a,x) \approx {50\over 3} \, 
\left({3+n\over 3}\right)^{-0.4} \, \bar\xi_{\rm L}(a,l)^{3/2}
\label{paddy}
\end{equation}
where,
\begin{equation}
l^3\approx x^3\bar\xi_{\rm NL}(a,x)\, \, ; \qquad
\bar\xi(x)={3\over x^3}\int_0^x {\rm d}y\ y^2 \, \xi(y)\, .
\label{suppl}
\end{equation}
Here, $\bar\xi(a,x)$ is the volume integral of the correlation
function over a sphere of comoving radius $x = r/a$ (the subscripts
refer to the non-linear and the linear regimes, respectively).  Since
$\bar\xi_{\rm L}(a,l)\propto l^{-(n+3)}$, it is easy to see that the
correlation function in equation~(\ref{xips}) has the same scaling as
equation~(\ref{paddy}).  

The predicted amplitudes of the non-linear correlation function 
(equations~\ref{xips} and~\ref{paddy}) can be compared after some
straightforward simplifications. For the initial spectrum 
$P(k)=A\,k^n$, the expression in equation~(\ref{xips}) can be 
written as 
\begin{equation}
\xi_{\rm sc}(a,r)= A \ N_{\rm sc}(n) \ a^{6\over (5+n)}\ r^{-\gamma}\, ,
\label{xips2}
\end{equation}
where $N_{\rm sc}$ contains all the normalization factors. Also,
equation (\ref{paddy}) simplifies to,
\begin{equation}
\xi_{\rm jmw}(a,r)= A \ N_{\rm jmw}(n) \ a^{6\over (5+n)}\ r^{-\gamma}\, .
\label{paddy2}\end{equation}
So, the task of comparing the amplitude of $\xi$
to the $N$-body calibrated formula simplifies to a comparison
of the normalization constants $N_{\rm sc}(n)$ and $N_{\rm jmw}(n)$. 
The results for the cases $n=0, -1,$ and $-2$ are shown in Table~1. 
Note that, for the $n=-2$ case, the agreement between our prediction and 
the $N$-body data is better than indicated in the table because, as noted
by JMW, their formula underestimates the non-linear amplitude of $\xi$ 
for this spectrum.

We can also compare the amplitude predicted by
equation~(\ref{xips}) directly with that measured in the $N$ body simulations 
of Efstathiou \etal (1988).  Their Fig.~4 shows ${\rm log}_{10} \Xi$ where 
$\Xi = (x/L)^\gamma \xi(x/L)$ for $-2\le n\le 1$.  At late times, so that 
discreteness effects are less important, the amplitudes of their correlation 
functions scale as $a^{6/(5+n)}$.  This scaling is consistent with that of 
equation~(\ref{xips}).  At the final output time in their simulations 
${\rm log}_{10} \Xi = -1.6, -1.4,$ and $-0.9$ for $n=1,0,$ and $-1$, 
respectively.  If we set $\Delta_{\rm nl}=178$ and use the fact that, 
in their simulations, $\bar\rho = 32^3/(aL)^3$ and 
$M_* = C_n\,a^{6/(3+n)}$, where $C_n=0.8, 0.71,$ and 0.53 (their eq.~15), 
then equation~(\ref{xips}) predicts that 
${\rm log}_{10} \Xi = -1.6, -1.3,$ and $-0.8$, respectively.  
Thus, our predicted amplitudes (equation~\ref{xips}) and those measured 
in the JMW and the Efstathiou \etal simulations agree to within about $20\%$. 
Some of this discrepancy is probably due to the fact that the 
Press--Schechter mass functions provide good, but not perfect, fits to the 
halo size distribution in the simulations.  We conclude that 
equation~(\ref{xips}) provides a good description of the shape and amplitude 
of the correlation function in the highly non-linear regime.

\begin{table}
\caption{Comparison of the amplitude of the non-linear $\xi$}
\medskip
\begin{tabular}{|c|c|c|c|}			\hline
$n$   & $N_{\rm sc}$ & $N_{\rm jmw}$  & Fractional ``Error'' \\ 
\hline
$0$   & $3.5$    & $3.1$      & 13 \% \\
$-1$  & $5.1$    & $5.0$      & 2 \%  \\
$-2$  & $17.4$   & $13.7$     & 27 \% \\
\hline
\end{tabular}
\label{Table}
\end{table}

\section{Relation to secondary infall models of spherical collapse}

The calculation of the previous section assumed that Press--Schechter 
halos are spherical, with density profiles having slope 
$\epsilon = 3(4+n)/(5+n)$.  That is, the previous section provided a 
relation between the initial fluctuation spectrum and the density 
profiles of non-linear, spherical, virialized halos (equation~\ref{sc}).  
It also showed that the correlation function that corresponds to this 
choice for the relation between $\epsilon$ and $n$ is in good agreement 
with that measured in $N$-body simulations.  However, this relation 
between $\epsilon$ and $n$ can be tested more directly.  

Recently, Crone \etal (1994) measured density profiles of virialized 
halos in their initially scale free simulations.  Although they did not 
compare their measurements with the relation of equation~(\ref{sc}), 
their simulations show that it is remarkably accurate.  Therefore, it is 
interesting to see if this relation can be derived directly from models of 
spherical collapse, rather than from the combination of the Press--Schechter 
model with the stable clustering hypothesis used in the previous section.
 
It is well known that if the initial density profile of a halo is a 
power law in radius (a `cone-hat', rather than a `tophat'), then the 
density profile of the collapsed halo is also a power law.  If the 
initial density profile is $\rho\propto r^{-\alpha}$, then the collapsed 
halo has slope
\begin{equation}
\epsilon = {3\alpha\over 1+\alpha} \qquad{\rm if\ }\alpha\!<\!2,
\label{slope}
\end{equation}
and $\epsilon = -2$ if $\alpha<2$ (Fillmore \& Goldreich 1984).  If one 
allows for non-radial orbits, then the restriction on $\alpha$ can be 
relaxed (White \& Zaritsky 1992).  This solution describes the shape 
of a single collapsed halo.  If, instead, we start with a distribution 
of halos identified in an initially Gaussian random field, then the 
question is: What is the initial density profile of a typical halo?  
Once we know the answer, we can set this value equal to $\alpha$ in 
equation~(\ref{slope}), and so calculate the typical density profile 
of collapsed halos.  

Hoffman \& Shaham (1985) argue that the initial progenitor halos may 
be related to peaks in the initial density field.  The density profiles 
of very high peaks have the same $r$ dependence as the correlation 
function of the underlying field (Section VII in Bardeen \etal 1986).  
Since the correlation function of the initial (scale free) field is 
$\propto r^{-(3+n)}$, Hoffman \& Shaham set $\alpha = 3+n$ in 
equation~(\ref{slope}) to obtain $\epsilon = 3(3+n)/(4+n)$.  This 
dependence on $n$ is different from that in equation~(\ref{sc}).  
However, it is not clear that the Hoffman--Shaham value for $\alpha$ is 
the relevant one.  This is because, as the peak height decreases, the 
peak profile becomes steeper than the correlation function (Fig.~8 in 
Bardeen \etal 1986).  As the universe expands, peaks of smaller and 
smaller height are able to collapse, so it is important to account for 
this difference between the average peak profiles and the shape of the 
initial correlation function.    

The mean peak profile, averaged over all curvatures and orientations, 
depends, to a good approximation, on a sum that involves the initial 
correlation function and its second derivative (equation~7.10 in 
Bardeen \etal 1986).  For initially scale free fields, the correlation 
function is $\propto r^{-(3+n)}$ so that its second derivative is 
$\propto r^{-(5+n)}$.  Thus, setting $\alpha = 4+n$ should give 
some indication of the shapes of collapsed peaks, when the initial 
peaks are not arbitarily high.  This value for $\alpha$ in 
equation~(\ref{slope}) gives $\epsilon = 3(4+n)/(5+n)$ for 
$-2\le n\le 1$.  While it is interesting that this relation between 
the non-linear density profile and the initial fluctuation field is 
the same as that obtained in the previous section using the stable 
clustering hypothesis, one would prefer a more rigorous derivation.  

\section{The onset of stable clustering}

Section 2 used the fact that the shape of the correlation 
function in the non-linear, stable clustering regime is related to the 
average density profile and mass distribution of virialized halos.  
It did not address the question of the scale on which stable clustering 
becomes a good approximation.  The usual assumption is that stable 
clustering is accurate when the density contrast is on the order of 
that required for virialization:  $\Delta_{\rm nl}\sim 200$.  On the 
other hand, $N$-body simlulations show that the onset of stable 
clustering occurs at higher density contrasts as $n$, the slope of the 
power spectrum of initial fluctuations, increases (Jain 1995).  This 
section uses the Press--Schechter description of clustering to show 
why this might be expected.  

Essentially, this section is motivated by the fact that not all 
Press--Schechter halos at the epoch $t_1$ evolve to become cores of 
halos by the epoch $t_2\!>\!t_1$.  For example, one might reasonably 
expect that for a halo of mass $M_2$ at the epoch $t_2$, the most massive 
subhalo $M_1\!<\!M_2$ at the epoch $t_1$ becomes the core.  If the density 
profile of $M_2$ at $t_2$ has the slope obtained in the previous section, 
then some or all of the other subhalos that make up the mass $(M_2-M_1)$ 
may have been destroyed as they merged to form the final halo.  This 
disruption of subhalos as they merge to form larger halos has been 
seen in $N$-body simulations (e.g. Fig.~2 in Efstathiou \etal 1988).  
These disrupted subhalos will not have evolved in accordance with the 
stable clustering hypothesis.  So, only the (most massive?) progenitor 
subhalo in the core is likely to have evolved consistently with both 
Press--Schechter and stable clustering.  This has the following consequence.  

It is usual to assume that stable clustering should be a good approximation 
on scales that are comparable to that of a typical Press--Schechter halo.  
This corresponds to scales on which the density is on the order of 
$\sim 200$ times the background density.  However, the discussion above 
suggests that the objects that are not in the core of a Press--Schechter 
halo may not satisfy the stable clustering hypothesis.  Therefore, the 
relevant density contrast for the onset of stable clustering is that 
associated with the core of the halo, rather than of the halo itself.  
Suppose we assume that the core size is the scale associated with the 
typical size of the most massive progenitor of a given halo.  Since 
the core is smaller (and more dense) than the halo itself, one might 
reasonably expect that stable clustering will only be a good approximation 
once the density contrast is somewhat larger than the canonical value of 
$\sim 200$ or so.  

To be more quantitative, consider a halo with mass $M_0$ at time $t_0$, 
and assume that a stable core was formed in its center when one of its 
progenitor subhalos had a mass of $M_0/2$.  To estimate the density of 
this half-mass core, we must estimate the scale associated with this core, 
which, by the stable clustering assumption, is the same physical scale 
it had at the epoch at which it virialized.  So, we must estimate the 
earliest epoch at which half the mass of the halo was first assembled 
into a virialized subhalo of mass $M_0/2$.  There are two ways to do 
this.  The first is to use the fact that 
\begin{equation}
P(M_1\!>\!M_0/2,t_1|M_0,t_0) = 
{\rm erfc}\,\left({\bar\omega\over \sqrt{2}} \right),
\end{equation}
where $P(M_1\!>\!M_0/2,t_1|M_0,t_0)$ denotes the probability that a mass 
element of $M_0$ was previously in an object of mass $M_1>M_0/2$ at the 
epoch $t_1$, and where
\begin{equation} 
\bar\omega = {\delta(t_1)-\delta(t_0)\over 
\sqrt{\left(\sigma_0^2(M_0/2)-\sigma_0^2(M_0)\right)}},
\end{equation}
where $\delta(t)$ describes the (linear theory) overdensity required for 
collapse at time $t$ (it decreases as the inverse of the expansion factor), 
and $\sigma_0^2(M)$ is the variance of the linear density field when 
smoothed with a filter containing mass $M$ (see Lacey \& Cole 1993 
for details).  Then, setting $P=1/2$ gives an estimate of 
$\delta(t_1)\equiv\delta(t_{\rm f})$ at the time $t_{\rm f}$ of formation.  
Since ${\rm erfc}\,(x) = 1-{\rm erf}\,(x) = 1/2$ when $x\approx 0.4769$, 
this means that $\bar\omega_{\rm f} = 0.4769\,\sqrt{2} = 0.67$.  This way
of estimating the halo formation time has been used by 
Navarro, Frenk \& White (1995).

However, as Lacey \& Cole (1993) note (and their Fig.~7 shows), this method 
provides a biased estimate of the formation time.  They show that a more 
precise estimate of the formation time is obtained by noting that, with 
this (half-mass assembled) definition of formation, different halos, each 
of mass $M_0$, may have formed at different times.  So, one should compute 
the distribution of formation epochs, and then compute the mean, or the 
most probable value, of this distribution.    
For initially scale free fields, the most probable value of the formation 
epoch, $t_{\rm f}$, is given by 
\begin{equation}
1 + z_{\rm f} = \left({t_0\over t_{\rm f}}\right)^{2/3} = 
1 + W_{\rm f}\ \sqrt{2^\beta - 1}\ \left({M_*\over M_0}\right)^{\beta/2} ,
\label{zform}
\end{equation}
where $\beta = (n+3)/3$, and $W_{\rm f}\approx 0.75$ (see Fig.~7 and 
equation~2.32 in Lacey \& Cole 1993).  This estimate differs  
slightly from the estimate of the previous paragraph, for which 
$W_{\rm f}\approx 0.67$.  

\begin{figure}
\centering
\mbox{\psfig{figure=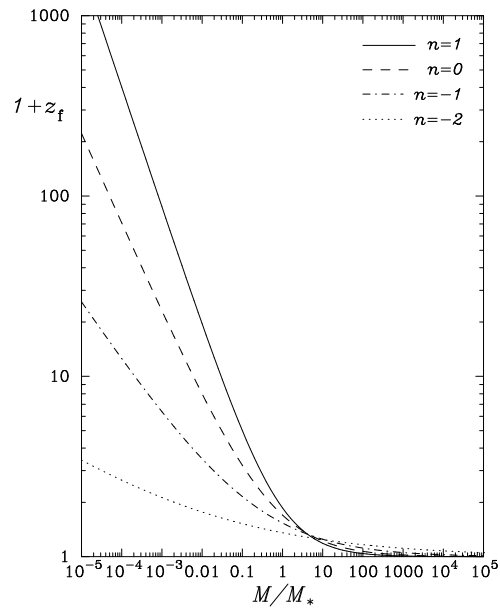,height=8cm,bbllx=79pt,bblly=125pt,bburx=489pt,bbury=631pt}}
\caption{
The epoch of formation $z_{\rm f}$ for halos with mass $M/M_*$, for 
various choices of $n$. }
\label{w07}
\end{figure}

Equation~(\ref{zform}) shows that the formation epoch is mass dependent.  
Fig.~\ref{w07} shows this distribution of formation epochs for $-2\le n\le 1$, 
with $W_{\rm f}= 0.7$ (the curves for $W_{\rm f}=0.67$ and 0.75 are very 
similar to the one shown).  Two features are obvious.  Halos that are 
less massive than about $10M_*$ or so form earlier (at higher $z_{\rm f}$) 
for large $n$ than for small $n$.  On the other hand, halos that are more 
massive than this scale with $n$ in the opposite sense.  It is easy to see 
that both these features are a consequence of two facts.  Namely, the 
distribution of Press--Schechter halo sizes becomes broader as $n$ 
decreases; when $n=1$ most halos have the characteristic mass $M_*$, 
but when $n=-2$, many halos have significantly different masses.  So, 
if $n=1$ initially, then halos that are significantly more massive than 
the characteristic mass at, say, the present time must have formed very 
recently (i.e., at low redshift), since they are extremely unlikely to 
have existed at early times when the characteristic mass was lower.  In 
contrast, if $n<-1$ initially, then some of these more massive halos 
may easily have been formed at earlier times (higher redshifts).  
Furthermore, the growth rate of the characteristic mass is more rapid 
for small values of $n$ (equation~\ref{mstar}).  Therefore, the 
asymptotic power law behaviour shown in the figure (for small $M/M_*$ 
as a function of $1+z_{\rm f}$) is a direct consequence of 
equation~(\ref{mstar}).

So, the way in which the epoch of formation of cores scales with $n$ 
depends on the ratio $M/M_*$.  However, most of the mass is in halos 
with masses $M\le M_*$ (the actual fraction is ${\rm erf}\,(1)=0.84$).  
This means that, for large $n$, most of the stable cores will virialize 
at earlier times than for lower $n$.  By the Press--Schechter hypothesis, 
progenitor cores will have had an average density of $\sim 200$ times the 
background density when they virialized (at $z_{\rm f}$).  This sets the 
scale of the halo.  Stable clustering implies that, in physical coordinates, 
the size of the core does not change, so that, by $t_0$, its comoving 
density will have increased by the cube of the expansion factor since 
the time $t_{\rm f}$ when it virialized.  So, on average, the average 
density of stable cores of Press--Schechter halos increases as $n$ 
increases.  Since stable clustering is only a good approximation in 
these cores, the density constrast on which stable clustering becomes 
a good approximation increases as $n$ increases.  

We can estimate the amplitude of the correlation function at which 
stable clustering becomes accurate as follows.  Assume that, when averaged
over all halos, the onset of stable clustering occurs at the half-mass 
radius of a halo with mass $M=f^3\,M_*$, where $f$ was given at the end 
of section 2.2.  Recall that this definition of $f$ was chosen so that, if 
all halos had this same mass, rather than the Press--Schechter distribution 
of masses, then the $r\to 0$ amplitude of the correlation function would be 
unchanged.  The ratio of this half-mass radius to the virial radius of such 
a halo is $0.5^{1/(3-\epsilon)}$.  On this scale, the correlation function is  
$\xi\, \gsim \,\lambda_n\,(4\pi\Delta_{\rm nl}/3)\ 2^{\gamma/(3-\epsilon)}$ 
which implies $\xi\, \gsim \,300$ for $n=-2$ and $\xi\, \gsim \,600$ for 
$n=0$.  This increase of the requisite amplitude of $\xi$ as $n$ increases 
is in qualitative agreement with recent measurements of the onset of stable 
clustering in $N$-body simulations for which $\xi\, \gsim \,200$ for 
$n=-2$ and $\xi\, \gsim \,1000$ for $n=0$ (Jain 1995).

\section{Discussion}

We have shown that requiring consistency between the stable clustering 
hypothesis and the Press--Schechter approach suggests that the density 
profiles of Press--Schechter halos should be power laws with slope 
$\epsilon = 3(4+n)/(5+n)$, where $n$ is the slope of the initial 
fluctuation spectrum (equation~\ref{sc}).  This value differs from 
that expected from secondary infall studies of the collapse of peaks 
associated with initially Gaussian fields (Hoffman \& Shaham 1985).  
The reasons for this difference were discussed in Section~3.  

The Press--Schechter model, with the stable clustering density profile, 
was used to compute the amplitude of the correlation function in the 
highly non-linear regime (Section 2).  This amplitude was shown to be 
in good quantitative agreement with that measured in $N$-body 
simulations.  Moreover, the amplitude was shown to be a function of 
the slope $n$ of the initial power spectrum.  The $n$-dependence 
resembles that which was obtained recently by Jain \etal (1995) 
and Padmanabhan \etal (1995).  

Our calculation of $\xi$ (equation~\ref{mcsilk}) rests on the 
assumption that, in this (small separation) regime, the correlation 
function is determined primarily by the density profiles of virialized 
halos.  That is, the crucial assumption is that, for these small 
separations, most pairs are most likely to be drawn from the same 
halo, so that correlations between different halos have a negligible 
effect on the shape of $\xi$.  While this assumption makes intuitive 
sense, we have yet to show that it is accurate.  We can do this as 
follows.  

Equation~(\ref{mclps}) shows that the correlation function can be written 
as a sum over halos of different masses.  To demonstrate that our approach 
is at least self-consistent, we must show that the correlation function is, 
indeed, determined primarily by halos having diameters that are larger 
than the separation scale $r$.  So, consider the limit $r\to 0$ of 
vanishingly small separations.  Then the amplitude of the correlation 
function is given by equation~(\ref{xips}).  Of course, in this limit, 
all halos are larger then the separation scale.  However, as a crude 
approximation that should also give some indication of the result on 
slightly larger scales, we will compute the fractional contribution to 
$\xi$ from `small' halos, which we will define to be all those with 
$M/M_* \le 0.2$.  Equation~(\ref{xips}) shows that this contribution is 
simply an incomplete Gamma function.  Thus, the small mass contribution is 
\begin{displaymath}
\gamma\Bigl[(11+n)/(10+2n),0.2^{n+3/3}\Bigr]\,\Big/\,
\Gamma\Bigl[(11+n)/(10+2n)\Bigr] ,
\end{displaymath}
which is 11, 15, 20, and 24 per cent for $n=1,0,-1$ and $-2$ respectively.  
For non-zero values of $r$, the corresponding integrals can be done 
numerically.  They also show that $\xi$ is determined primarily by pairs 
within larger halos, so that our neglect of halo--halo correlations 
is justified.  

The other crucial ingredient in our calculation is knowledge of the 
density profiles of virialized halos.  We assumed that these profiles 
were power laws, and determined the slopes of the power laws by requiring 
that they yield a correlation function with the shape required by stable 
clustering.  However, recent work suggests that, while the profile shape 
may be described by an average power law slope that is consistent with 
equation~(\ref{sc}), the density profiles of halos that form in $N$-body 
simulations of gravitational clustering from scale free initial conditions 
are not simple power laws (Navarro, Frenk \& White 1995; Cole \& Lacey 1995; 
Tormen 1995).  Rather, the density profiles seem to be well fitted by a 
function of the form $\rho \propto 1/[r(r+b)^2]$, where the core radius 
$b$ depends both on the mass $M$ and the shape of the initial power 
spectrum.  

\begin{figure}
\centering
\mbox{\psfig{figure=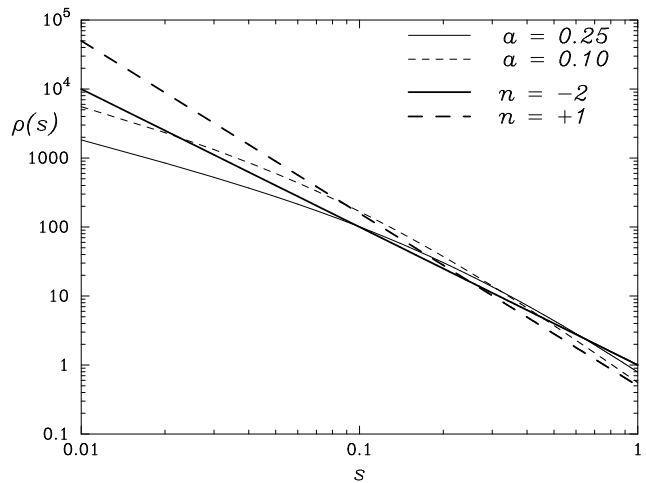,height=2.5in,bbllx=54pt,bblly=70pt,bburx=563pt,bbury=449pt}}
\caption{
The density profile $\rho(s) \propto 1/[s(s+a)^2]$, 
where $s=r/r_{\rm vir}$, $a = b/r_{\rm vir}$, and $b$ is the core radius, 
for a range of values of $a$.  When $s \gsim 0.1$, the power law profile
that corresponds to $n=-2$ (solid bold curve) describes the $a=0.25$ profile 
quite well.  The dashed bold curve shows the power law corresponding to 
the $n=1$ case; when $s \gsim 0.1$ it fits the $a=0.1$ profile reasonably 
well.}
\label{nfw}
\end{figure}

We expect our formalism to work for these non-power-law profiles 
also because, except for the inner regions $r\sim 0.1\,r_{\rm vir}$, where
$r_{\rm vir}$ is the virial radius, they are reasonably well described by 
the power-law profiles we have been considering.  To illustrate this, 
Fig.~\ref{nfw} shows the non-power law profile $\rho(s) \propto 1/[s(s+a)^2]$, 
where $s=r/r_{\rm vir}$ and the core radius is given by $a = b/r_{\rm vir}$.  
The plot shows this profile for a range of values of the scaled core 
radius $a$.  
These values were chosen because $N$-body simulations show that 
$a\sim 0.25$ describes the $n=-2$ simulations quite well, whereas $a$ 
is smaller ($\sim 0.15$) when $n=0$ (Cole \& Lacey 1995).  For comparison, 
the thicker curves show the corresponding stable clustering power law 
profiles for the two extreme cases studied here, $n=-2$ and $n=1$.  Notice 
that, in the outer regions where $s \gsim 0.1$, the $n=-2$ (solid bold) 
curve describes the $a=0.25$ profile quite well, whereas the $n=1$ 
(dashed bold) curve fits the $a=0.1$ profile.  

Now, the correlation function in the non-linear regime is essentially an 
integral over the density profile (to give $\lambda_M$), followed by an 
integral over the mass function.  
For the choices of $n$ and $a$ shown in Fig.~1,
the corresponding $\lambda_M(r)$ curves for the power law and the non-power 
law cases are similar, on scales larger than about $0.1\,r_{\rm vir}$.  This
is hardly surprising, since Fig.~\ref{nfw} shows that the Navarro \etal 
and power law density profiles $\rho(s)$ are similar in the outer regions 
where $s \gsim 0.1$ (which contain most of the mass of a halo), and 
$\lambda_M$ is simply the density profile convolved with itself.  

Since, in 
our formalism, $\xi$ is simply the sum of many $\lambda_M$ curves, this
suggests that $\xi$ for these non-power law profiles should be similar to
that for the power law profiles considered in Section~2, at least for 
separations that are on the order of about $0.1\, r_*$ and larger.  We have 
verified that this is indeed the case: for scales on which $\xi(r)\lsim 1000$, 
the predicted amplitude of $\xi$ changes by less than about $10\%$ for 
$-2<n<1$, if we use the Navarro \etal profile (with the values of $a$ shown
in Fig.~1) instead of power law profiles.  

On scales smaller than $\sim 0.1\, r_*$, it is necessary to take into 
account the dependence of the core radius on the mass of the halo.  
Simulations show that, typically, $a$ decreases as $M$ decreases.  
Navarro \etal (1995) argue that the qualitative form of 
this relation can be understood in terms of the formation times of 
Press--Schechter halos.  Including this trend improves the agreement of 
our predicted $\xi$ with the result from the Navarro et al profile. 
The asymptotic behavior of $\xi(r)$ as $r\to 0$ however, is more subtle, 
as it is 
very sensitive to the $M \to 0$ behavior of the core radius $a$ and the
Press-Schechter $n(M)$. Here we simply note that by requiring
that $\xi(r)$ have the stable clustering shape, we can
obtain an independent constraint on the relation between 
$a$, $M$, and $n$ for these non-power law profiles. 

Since all of the results of this paper concern the stable clustering 
regime, Section~4 discussed the scale on which stable clustering should 
become a good approximation.  It used the Lacey \& Cole (1993) analysis 
of the merger histories of Press--Schechter halos to show that the onset 
of stable clustering occurs at higher density contrasts as the slope of 
the initial fluctuation power spectrum $n$ increases.  This is in 
qualitative agreement with $N$-body simulations (Jain 1995).   

Our derivation of the amplitude of the correlation function 
in the stable clustering regime, using the Press--Schechter mass 
multiplicity function, can be extended to cosmological models in 
which $\Omega\le 1$.  This is because the effect of $\Omega$ on the
density profiles of halos has been calculated (e.g. Hoffman \& 
Shaham 1985; Hoffman 1988; White \& Zaritsky 1992), as has the effect 
on the evolution of the Press--Schechter mass function 
(Lacey \& Cole 1993).  Furthermore, recall that both the density profiles 
and the Press--Schechter multiplicity function depend strongly on the 
shape of the initial power spectrum.  Thus, our calculation of $\xi$ 
shows explicitly how the non-linear correlation function depends on the 
shape and amplitude of the linear power spectrum.  Therefore, if our 
approach is correct, then by requiring consistency between the shape 
of the mass multiplicity function and the shape of the non-linear 
correlation function, one can estimate the shape and amplitude of the 
initial perturbation spectrum.  

\section*{Acknowledgments}
We thank Gerard Lemson for sending us copies of his insightful 
Ph.D. thesis, and Simon White and Saleem Zaroubi for many stimulating 
discussions and helpful suggestions.

\end{document}